\documentclass{PoS}
\usepackage{amsmath}
\usepackage{multirow}
\usepackage{slashed}
\usepackage{epsfig,psfrag,rotating,soul}
\usepackage[x11names,table]{xcolor}
\usepackage{hhline}
\usepackage{cite}
\newcommand{\nn}{\nonumber}
\newcommand{\mL}{\mathcal{L}}
\newcommand{\mO}{\mathcal{O}}

\def\significance{\sigma^{\rm stat}}
\title{Collider phenomenology of vector resonances in WZ scattering processes}
\ShortTitle{Collider phenomenology of vector resonances}
\author{\speaker{Rafael L. Delgado},$^a$ A. Dobado,$^b$ D. Espriu,$^c$ C. Garcia-Garcia,$^d$ M.J. Herrero,$^d$ X.~Marcano$^e$ and J.J. Sanz-Cillero$^b$\vspace{.2cm}\\
  \llap{$^a$}Physik-Department T30f, Technische Universit\"at M\"unchen,\\ 
  James-Franck-Str. 1, D-85747 Garching, Germany\\
  \llap{$^b$}Departamento de F\'{\i}sica Te\'orica, Universidad Complutense de Madrid,\\
  Plaza de las Ciencias 1, 28040 Madrid, Spain\\
  \llap{$^c$}Departament de F\'{\i}sica Qu\`antica i Astrof\'{\i}sica and Institut de Ci\`encies del Cosmos (ICCUB),\\
  Universitat de Barcelona, Mart\'{\i} i Franqu\`es 1, 08028 Barcelona, Catalonia, Spain\\
  \llap{$^d$}Departamento de F\'{\i}sica Te\'orica and Instituto de F\'{\i}sica Te\'orica, IFT-UAM/CSIC,\\
  Universidad Aut\'onoma de Madrid, Cantoblanco, 28049 Madrid, Spain.\\
  \llap{$^e$}Laboratoire de Physique Th\'eorique CNRS,\\
  Universit\'e Paris-Sud, Universit\'e Paris-Saclay, 91405 Orsay, France\vspace{.2cm}\\
  E-mail: \email{rafael.delgado@tum.de}, \email{dobado@fis.ucm.es}, \email{espriu@icc.ub.edu}, \email{claudia.garcia@uam.es},
  \email{maria.herrero@uam.es}, \email{xabier.marcano@th.u-psud.fr}, \email{jjsanzcillero@ucm.es}}
\abstract{%
We study the production of vector resonances at the LHC via $WZ$ scattering processes and explore the sensitivities to these resonances for the expected future LHC luminosities. The electroweak chiral Lagrangian and the Inverse Amplitude Method (IAM) are used for analyzing a dynamically generated vector resonance, whose origin would be the (hypothetically strong) self interactions of the longitudinal gauge bosons, $W_L$ and $Z_L$. We implement the unitarized scattering amplitudes into a single model, the IAM-MC, that has been adapted to MadGraph~5. It is written in terms of the electroweak chiral Lagrangian and an additional effective Proca Lagrangian for the vector resonances, so that it reproduces the resonant behavior of the IAM and allows us to perform a realistic study of signal versus background at the LHC. We focus on the $pp\to WZjj$ channel, discussing first on the potential of the hadronic and semileptonic channels of the final $WZ$, and next exploring in more detail the clearest signals. These are provided by the leptonic decays of the gauge bosons, leading to a final state with $l^+_1l^-_1l^+_2\nu jj$, $l=e,\mu$, having a very distinctive signature, and showing clearly the emergence of the resonances with masses in the range of $1.5$-$2.5\,{\rm TeV}$, which we have explored.}
\FullConference{XIII Quark Confinement and the Hadron Spectrum - Confinement2018\\
		31 July - 6 August 2018\\
		Maynooth University, Ireland}
\begin{document}
\section{Introduction}
These proceedings are based on our work~\cite{Delgado:2017cls}. The ATLAS and CMS experiments at CERN discovered a new scalar boson with the properties of the Standard Model (SM) one~\cite{Aad:2012tfa,Chatrchyan:2012xdj}, and an energy gap for any new physics~\cite{Khachatryan:2014jba, Aaboud:2017eta, Sirunyan:2017acf}. The Higgs boson is a key component of the electroweak symmetry breaking sector (EWSBS) of the SM. Since, at the LHC, we are exploring through direct search the EWSBS, the new physics (if any) could lie on it. The existence of an energy gap between the electroweak scale and the new physics scale (if any) would naturally fit into a BSM model with a strongly interacting EWSBS. These models introduce a new energy scale $f\gg v = 246\,{\rm GeV}$ where some new strong interactions trigger the dynamical breaking of a global symmetry group $G$ to a certain subgroup $H$. As indicated by the Equivalence Theorem (ET), at high energies $\gg v$, the constituents of the EWSBS behave as scalar Goldstone bosons. At lower energies, 3 of these Goldstone bosons give rise to the longitudinal components of gauge bosons. Hence, taking into account the SM suppression of the longitudinal gauge boson production at the LHC, a resonance on the longitudinal gauge boson scattering processes would be a smoking gun for new physics at the LHC involving the EWSBS.

There are two approaches for studying the collider phenomenology of beyond-SM (BSM) physics. The first one, top-down, takes a particular model with a UV-completion scheme which is studied at the TeV scale. This model can be a fully renormalizable one. The disadvantage is that we have no clue about the actual UV-completion of the underlying BSM theory (if any) and some BSM models, like the MSSM, have $\sim 100$ free parameters.

The second approach, bottom-up, involves developing an effective field theory (EFT) as general as possible. In particular, without making assumptions about the particular UV-completion scheme. In this work, the second approach is used. Hence, we will assume the SM spontaneous symmetry breaking pattern $SU(2)_L\times SU(2)_R\to SU(2)_{L+R}$, which involves 3 Goldstone bosons and the Higgs boson, and is the minimum to generate the electroweak (EW) gauge bosons masses (longitudinal modes of $W^\pm,Z$) while preserving the custodial symmetry $SU(2)_C=SU(2)_{L+R}$ (SM tree level relation $m_W=\cos\theta_W m_Z$), by means of the Effective Electroweak Chiral Lagrangian (EChL). It was developed from the eighties~\cite{Appelquist:1980vg, Longhitano:1980iz, Longhitano:1980tm, Chanowitz:1985hj, Cheyette:1987jf, Dobado:1989ax, Dobado:1989ue}, alongside the well established chiral perturbation theory (ChPT) of low energy QCD~\cite{Weinberg:1978kz, Gasser:1984gg, Gasser:1983yg}. It was used in the early nineties for LEP phenomenology \cite{Dobado:1990zh, Espriu:1991vm}, and for LHC prospects (mostly Higgs)~\cite{Dobado:1990jy, Dobado:1990am, Dobado:1995qy, Dobado:1999xb}. Although in principle it described just the interactions among EW Goldstones, it has incorporated the scalar field $H$ in the last years as a consequence of the discovery of a light Higgs-like particle~\cite{Feruglio:1992wf, Alonso:2012px, Buchalla:2013rka, Espriu:2012ih, Delgado:2013loa, Delgado:2013hxa, Brivio:2013pma, Espriu:2013fia, Espriu:2014jya, Delgado:2014jda, Buchalla:2015qju, Arnan:2015csa}.

If the electroweak sector happens to be strongly interacting, the perturbative analysis will break at the TeV scale. As well known in the case of low energy QCD~\cite{Weinberg:1978kz, Gasser:1984gg, Gasser:1983yg}, dispersion relations, encoded in the so--called unitarization procedures, are needed. A detailed study regarding the usage of dispersion relations, including coupled channels, can be found on~\cite{Delgado:2015kxa}. This work, based on~\cite{Delgado:2014jda, Delgado:2017cls}, is part of the effort on exploring the main implication of the EChL for LHC phenomenology. The absence of signals of strongly interacting EWSBS sets experimental bounds on the values of the chiral parameters of the EChL~\cite{Falkowski:2013dza, Brivio:2013pma, Khachatryan:2014jba, Aad:2014zda, ATLAS:2014yka, Fabbrichesi:2015hsa, Buchalla:2015qju, Aaboud:2016uuk, deBlas:2018tjm}.

For studying its collider phenomenology, we introduce a unitarized EChL description of $WZ$ scattering on MadGraph~5 by means of an intermediate effective Proca Lagrangian. Other approaches found on the literature are form-factors~\cite{Arnold:2008rz} or modified Feynman vertices~\cite{Kilian:2014zja}. The direct output of dispersion relations is an on-shell matrix element, whereas the input of Monte Carlo programs are Feynman rules and, of course, they deal with off-shell processes.

\section{EChL and Effective Proca Lagrangian}
In this work, we compute the cross section for the processes $pp\to W^+Z jj\to l^+l^+l^-\nu jj$, where the vector resonance is produced in the intermediate VBS subprocess $WZ\to WZ$, by means of the IAM and an effective Proca Lagrangian. We use the non-linear EChL, $\mL = \mL_2 + \mL_4$, with a derivative expansion,\par\noindent%
{\scriptsize
  \begin{align}
	  \mL_2 &= -\frac{1}{2 g^2} {\rm Tr}\Big(\hat{W}_{\mu\nu}\hat{W}^{\mu\nu}\Big) -\frac{1}{2 g'^{2}} %
	    {\rm Tr}\Big(\hat{B}_{\mu\nu}\hat{B}^{\mu\nu}\Big) %
	    +\frac{v^2}{4}\left[1+2a\frac{H}{v}+b\frac{H^2}{v^2}\right] {\rm Tr} \Big(D^\mu U^\dagger D_\mu U \Big) %
	    +\frac{1}{2}\partial^\mu H\,\partial_\mu H + \dots\,\label{EChL2}\\
	  \mL_4 &= a_1 {\rm Tr}\Big( U \hat{B}_{\mu\nu} U^\dagger \hat{W}^{\mu\nu}\Big) %
	    + ia_2{\rm Tr}\Big(U\hat{B}_{\mu\nu} U^\dagger [{\cal V}^\mu, {\cal V}^\nu ]\Big) %
	    - ia_3{\rm Tr}\Big(\hat{W}_{\mu\nu}[{\cal V}^\mu, {\cal V}^\nu]\Big) %
	    +a_4\Big[{\rm Tr}({\cal V}_\mu {\cal V}_\nu) \Big] \Big[{\rm Tr}({\cal V}^\mu {\cal V}^\nu)\Big] \nn\\
	  &+a_5 \Big[{\rm Tr}({\cal V}_\mu {\cal V}^\mu)\Big] \Big[{\rm Tr}({\cal V}_\nu {\cal V}^\nu)\Big] %
	    -c_{W}\frac{H}{v} {\rm Tr}\Big(\hat{W}_{\mu\nu} \hat{W}^{\mu\nu}\Big) %
	    -c_B\frac{H}{v}\, {\rm Tr} \Big(\hat{B}_{\mu\nu} \hat{B}^{\mu\nu} \Big) + \dots
  \end{align}}%
Here, $U(w^\pm,z) = 1 + iw^a\tau^a/v +\mO(w^2)$, $D_\mu U = \partial_\mu U + i\hat{W}_\mu U -iU\hat{B}_\mu$, $\hat{W}_{\mu\nu} = \partial_\mu \hat{W}_\nu - \partial_\nu \hat{W}_\mu +i[\hat{W}_\mu,\hat{W}_\nu ],\;\hat{B}_{\mu\nu} = \partial_\mu \hat{B}_\nu -\partial_\nu \hat{B}_\mu$, $\hat{W}_\mu = g \vec{W}_\mu \vec{\tau}/2 ,\;\hat{B}_\mu = g'\, B_\mu \tau^3/2$ and ${\cal V}_\mu = (D_\mu U) U^\dagger$.

Note that some higher order operators that appear at dimension 8 in linear representation~\cite{Contino:2013kra, Alloul:2013naa} can contribute to a lower order in the non-linear one. In particular, this is the case of the chiral parameters $a_4$ and $a_5$, whose contribution is crucial for the processes we are considering here.

We use the on-shell Vector Boson Scattering (VBS) matrix elements computed by mean of the IAM~\cite{Delgado:2017cls} to adjust an effective Proca Lagrangian so that it reproduces the behaviour of the computed matrix elements up to the first resonance, which is the energy scale where the underlying EChL breaks. The Proca Lagrangian can be directly introduced inside MadGraph~5~\cite{Alwall:2014hca,Frederix:2018nkq} by means of FeynRules~\cite{Alloul:2013bka}. The key element is that we let the effective Proca couplings to be functions on the scale energy of the process, by means of \emph{ad hoc} Fortran functions inside our UFO model\footnote{Universal FeynRules Output, see Ref.~\cite{Degrande:2011ua}.}. In some way, this approach has some similarities with other approaches like the form-factor one~\cite{Arnold:2008rz} or with the effective approach of Kilian \emph{et. al.} (appendix~D of Ref.~\cite{Kilian:2014zja}). However, our underlying physical model is pretty different, specially from the form-factor approach, since we are, indeed, considering a BSM resonance coming from a strongly interacting EWSBS. This is why we are introducing an effective Proca Lagrangian, which explicitly introduces a vector resonance $V$ with a mass $M_V$, a width $\Gamma_V$ (that enters into the $V$ propagator) and couplings $f_V$ and $g_V$,\par\noindent%
{\scriptsize
  \begin{multline} %
	  \mL_V = 
	  \mL_V^{\rm kin} - \frac{if_V}{v^2} \bigg[m_W^2 V^0_\nu (W^+_\mu W^{-\,\mu\nu} -W^-_\mu W^{+\,\mu\nu}) %
	  + m_W m_Z V^+_\nu (W^-_\mu Z^{\mu\nu} -Z_\mu W^{-\,\mu\nu}) %
	  + m_W m_Z V^-_\nu (Z_\mu W^{+\, \mu\nu}- W^+_\mu Z^{\mu\nu}) \bigg] \\
	  +\frac{2ig_V}{v^2}\bigg[m_W^2 V^{0\,\,\mu\nu} W_\mu^+ W_\nu^- %
	  + m_W \, m_Z\, V^{+\,\, \mu\nu} W_\mu^-Z_\nu %
	  + m_W \, m_Z\, V^{-\,\, \mu\nu} Z_\mu W_\nu^+ \bigg], %
	  \label{LVugauge}
  \end{multline}}%
where $V^a_{\mu\nu}= \partial_\mu V^a_\nu - \partial_\nu V^a_\mu$ ($a=\pm,0$), $W^a_{\mu\nu}=\partial_\mu W^a_\nu - \partial_\nu W^a_\mu$ ($a=\pm$), and $Z_{\mu\nu}= \partial_\mu Z_\nu - \partial_\nu Z_\mu$. Our requirements are~\cite{Delgado:2017cls}:
\begin{itemize}
\item At low energies, below the resonance, the predictions from the effective Proca Lagrangian should mimic the unitarized scattering matrix element.
\item Above the resonance, the cross section should not grow faster than the Froissart bound. That is, $\sigma(s)\le\sigma_0\log^2\left(s/s_0\right)$.
\end{itemize}
Since we are studying deviations from the SM coming from the EWSBS, we are mostly focused on the longitudinal polarizations, so that we can set $f_V=0$ on Eq.~(\ref{LVugauge}). The function $g_V^2(z) = \left[\theta(M_V^2-s)(M_V^2/z) + \theta(s-M_V^2)(M_V^4/z^2)\right]\cdot g_V^2(M_V^2)$ is well suited for the coupling $g_V$, where $z=s,t,u$ when the resonance $V$ is propagating in the $s$, $t$ and $u$ channels, respectively.

Hence, for each benchmark point, we set $M_V$ and $\Gamma_V$ to the pole of the unitarized scattering amplitude. We extract $g_V(M_V^2)$ requiring that, for $s=M_V^2$ (on the resonance peak),  $\left\lvert a_{11}^{{\rm EChL}_{\rm tree}}+a_{11}^V\right\rvert = \left\lvert a_{11}^{\rm IAM}\right\rvert$, where $a_{11}$ stands for the isovector partial wave ($IJ=11$, see Ref.~\cite{Delgado:2015kxa}); ${\rm EChL}_{\rm tree}$, for the perturbative $\mL_2$ EChL amplitude [Eq.~(\ref{EChL2})]; $V$, for the Proca Lagrangian [Eq.~(\ref{LVugauge})]; and ${\rm IAM}$, for the unitarized scattering amplitude. Then, we substitute $g_V$ by $g_V(s)$, $g_V(t)$ and $g_V(u)$ when the resonance $V$ appears in the $s$, $t$ and $u$ channels, respectively.

\section{Results}\label{secresults}
\begin{figure}[th]
\null%
\hfill\includegraphics[width=.33\textwidth]{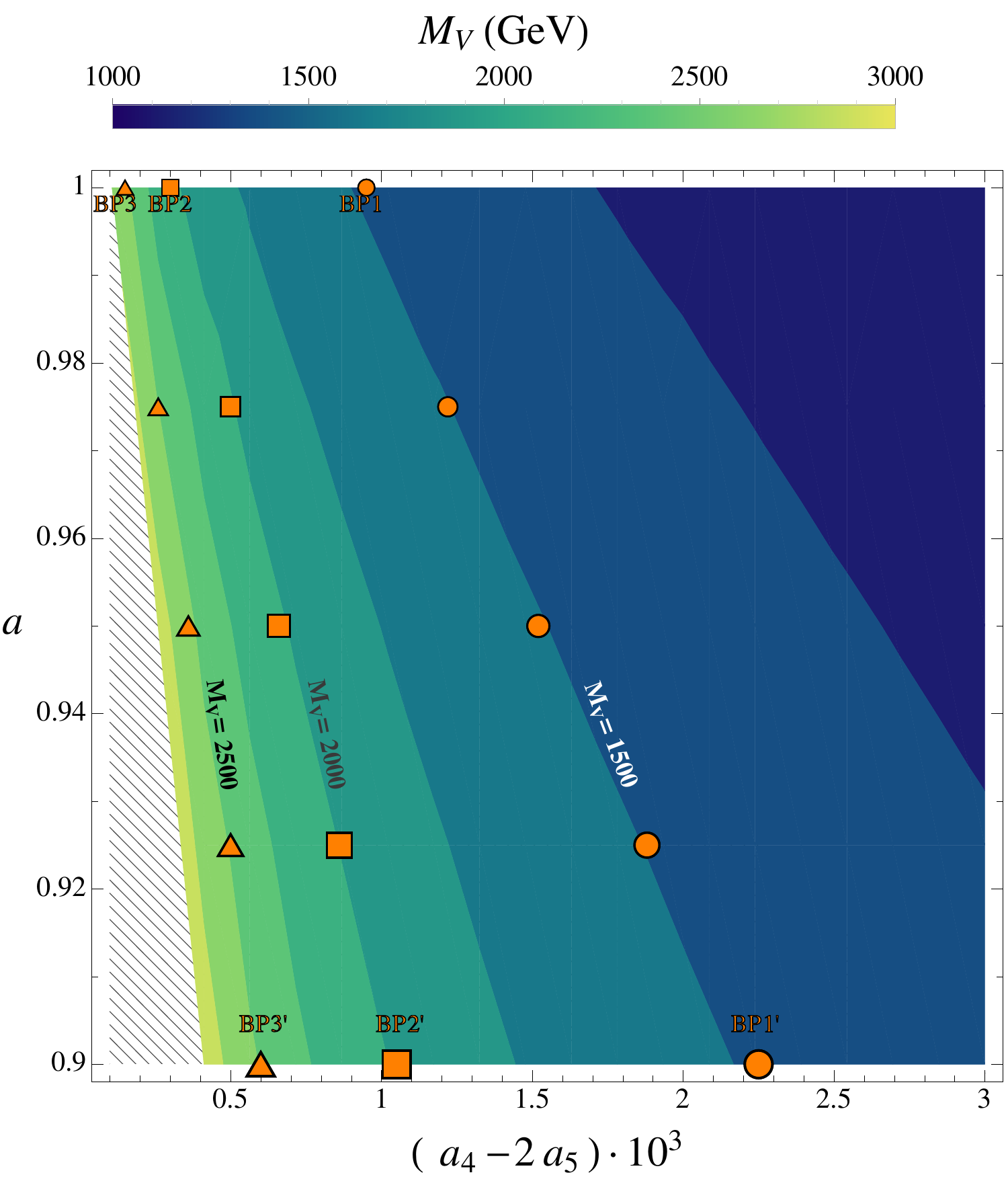}%
\hfill\includegraphics[width=.33\textwidth]{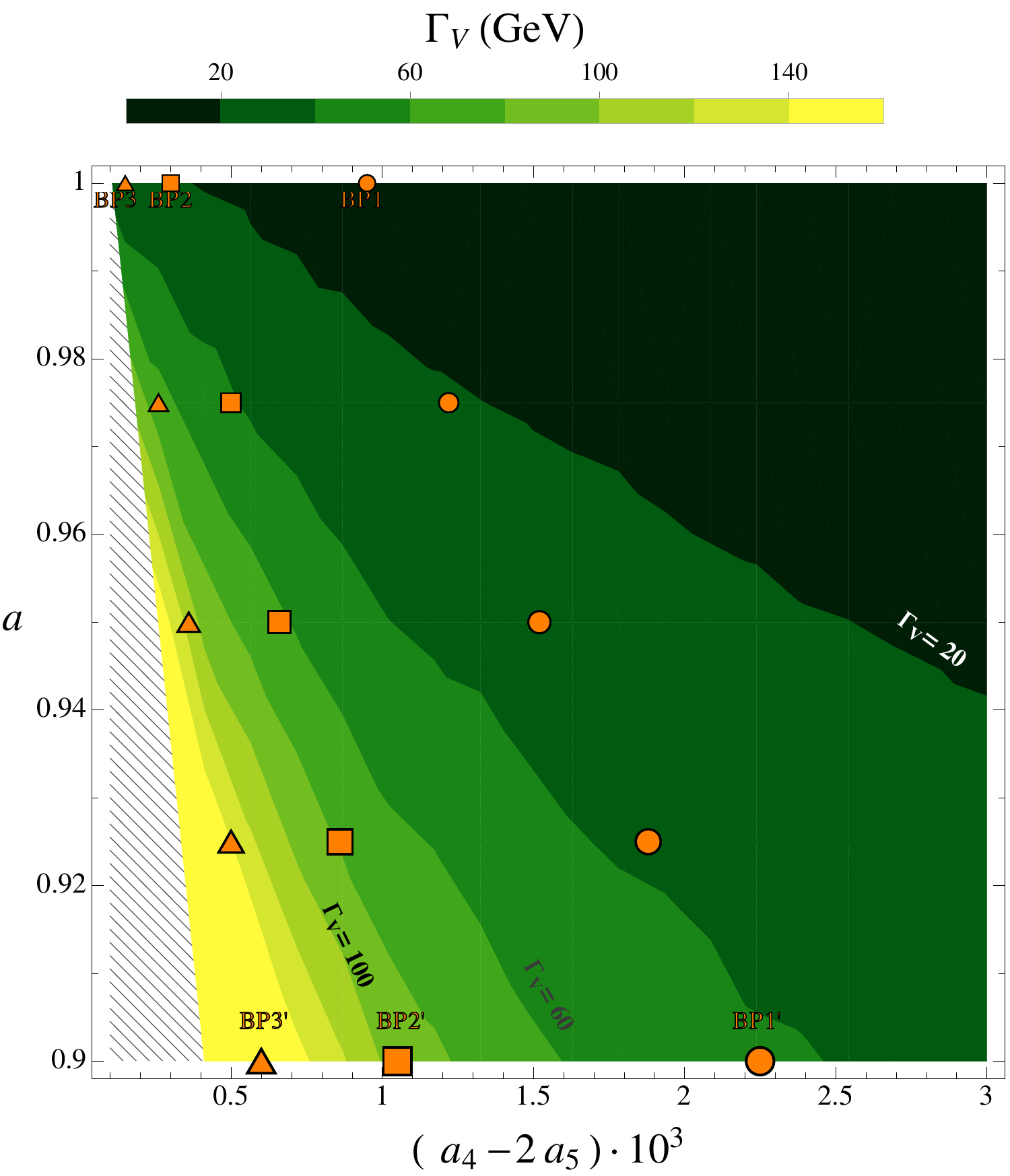}%
\hfill\includegraphics[width=.33\textwidth]{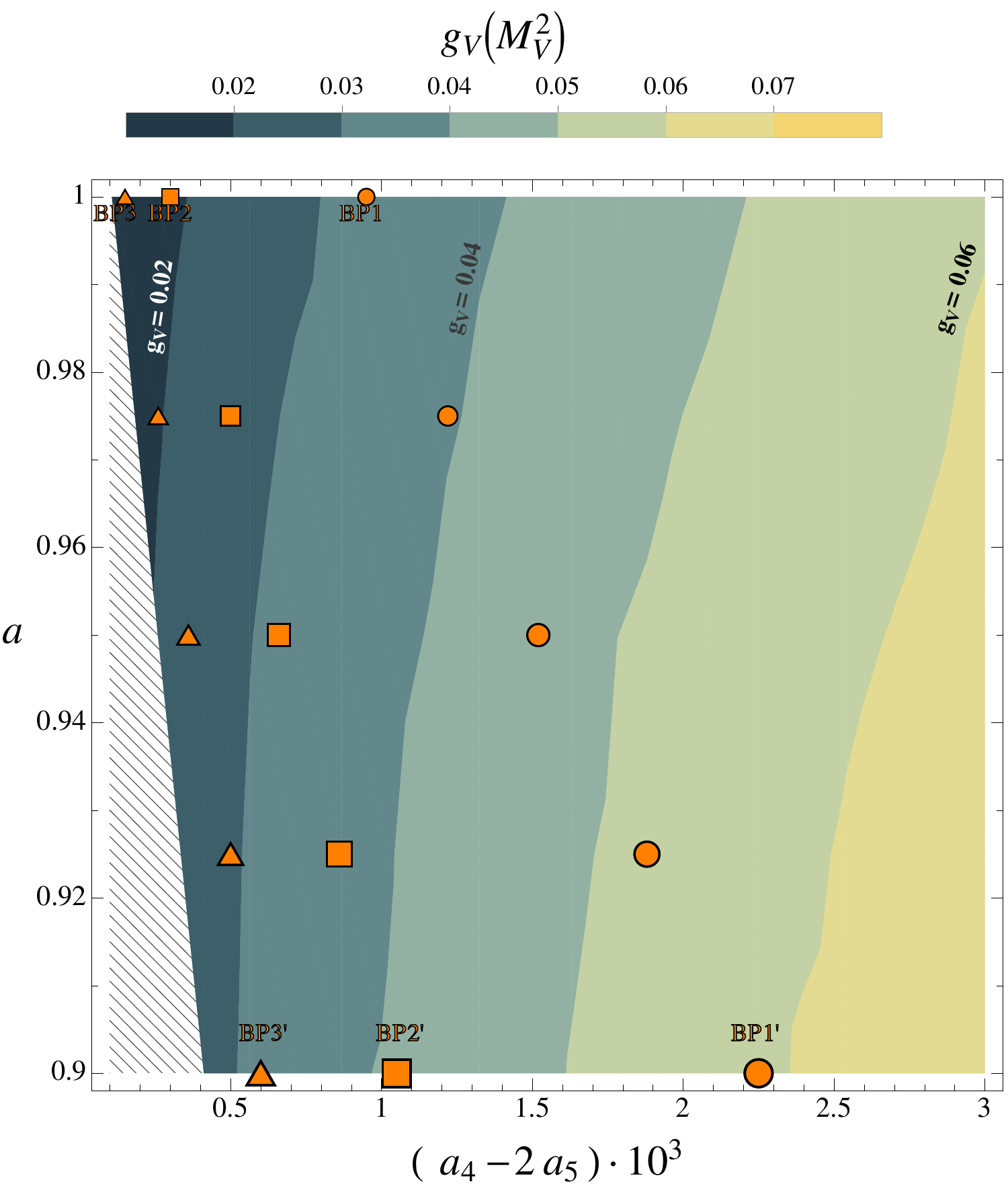}%
\hfill\null
\caption{\small Selected benchmark points, with $M_V$, $\Gamma_V$ and $g_V(M_V^2)$. From the IAM $IJ=11$ partial wave. Only the extremes $a=0.9$ and $a=1$ are used for the main analysis.}\label{BPfig}
\end{figure}

We have chosen 6 benchmark points (BPs), which are cited on table~\ref{BPtable} and Fig.~\ref{BPfig}. These are sets of $a$, $a_4$ and $a_5$ chiral parameters that have been chosen to dynamically generate resonances in the isovector $IJ=11$ channel, with masses around $1.5$, $2$ and $2.5\,{\rm TeV}$.

\begin{table}[t!h]
\begin{center}
\vspace{.2cm}
\begin{tabular}{ |c|c|c|c|c|c|c| }
\hline
\rule{0pt}{1ex}
{\footnotesize {\bf BP}} & {\footnotesize {\bf $M_V ({\rm GeV})$}}  & {\footnotesize  {\bf $\Gamma_V ({\rm GeV)}$}}  & {\footnotesize {\bf $g_V(M_V^2)$}}  & {\footnotesize {\bf $a$}}  & {\footnotesize {\bf $a_4 \cdot 10^{4}$}}  & {\footnotesize {\bf $a_5\cdot 10^{4}$}}
\\[5pt] \hline
\rule{0pt}{1ex}
BP1  & $\quad 1476 \quad $ & $\quad 14 \quad $ & $ \quad 0.033  \quad $ & $ \quad 1 \quad $ & $ \quad 3.5 \quad $ & $ \quad -3 \quad $
\\[5pt] \hline
\rule{0pt}{1ex}
BP2  & $\quad 2039 \quad $ & $\quad 21 \quad $ & $ \quad 0.018  \quad $ & $ \quad 1 \quad $ & $ \quad 1 \quad $ & $ \quad -1 \quad $
\\[5pt] \hline
\rule{0pt}{1ex}
BP3  & $\quad 2472 \quad $ & $\quad 27 \quad $ & $ \quad 0.013  \quad $ & $ \quad 1 \quad $ & $ \quad 0.5 \quad $ & $ \quad -0.5 \quad $
\\[5pt] \hline
\rule{0pt}{1ex}
BP1' & $\quad 1479 \quad $ & $\quad 42 \quad $ & $ \quad 0.058  \quad $ & $ \quad 0.9 \quad $ & $ \quad 9.5 \quad $ & $ \quad -6.5 \quad $
\\[5pt] \hline
\rule{0pt}{1ex}
BP2' & $\quad 1980 \quad $ & $\quad 97 \quad $ & $ \quad 0.042  \quad $ & $ \quad 0.9 \quad $ & $ \quad 5.5 \quad $ & $ \quad -2.5\quad $
\\[5pt] \hline
\rule{0pt}{1ex}
BP3' & $\quad 2480 \quad $ & $\quad 183 \quad $ & $ \quad 0.033  \quad $ & $ \quad 0.9 \quad $ & $ \quad 4\quad $ & $ \quad -1 \quad $
\\[5pt] \hline
\end{tabular}
\caption{\small Selected benchmark points (BPs) of dynamically generated vector resonances. The mass, $M_V$, width, $\Gamma_V$, coupling to gauge bosons, $g_V(M_V)$, and relevant chiral parameters, $a$, $a_4$ and $a_5$ are given for each of them. $b$ is fixed to $b=a^2$. This table is generated using the FORTRAN code that implements the EChL+IAM framework, borrowed from the authors in Refs.~\cite{Espriu:2012ih,Espriu:2013fia,Espriu:2014jya}.}\label{BPtable}
\end{center}
\end{table}

\begin{figure}[th]
\null%
\hfill\includegraphics[width=.49\textwidth]{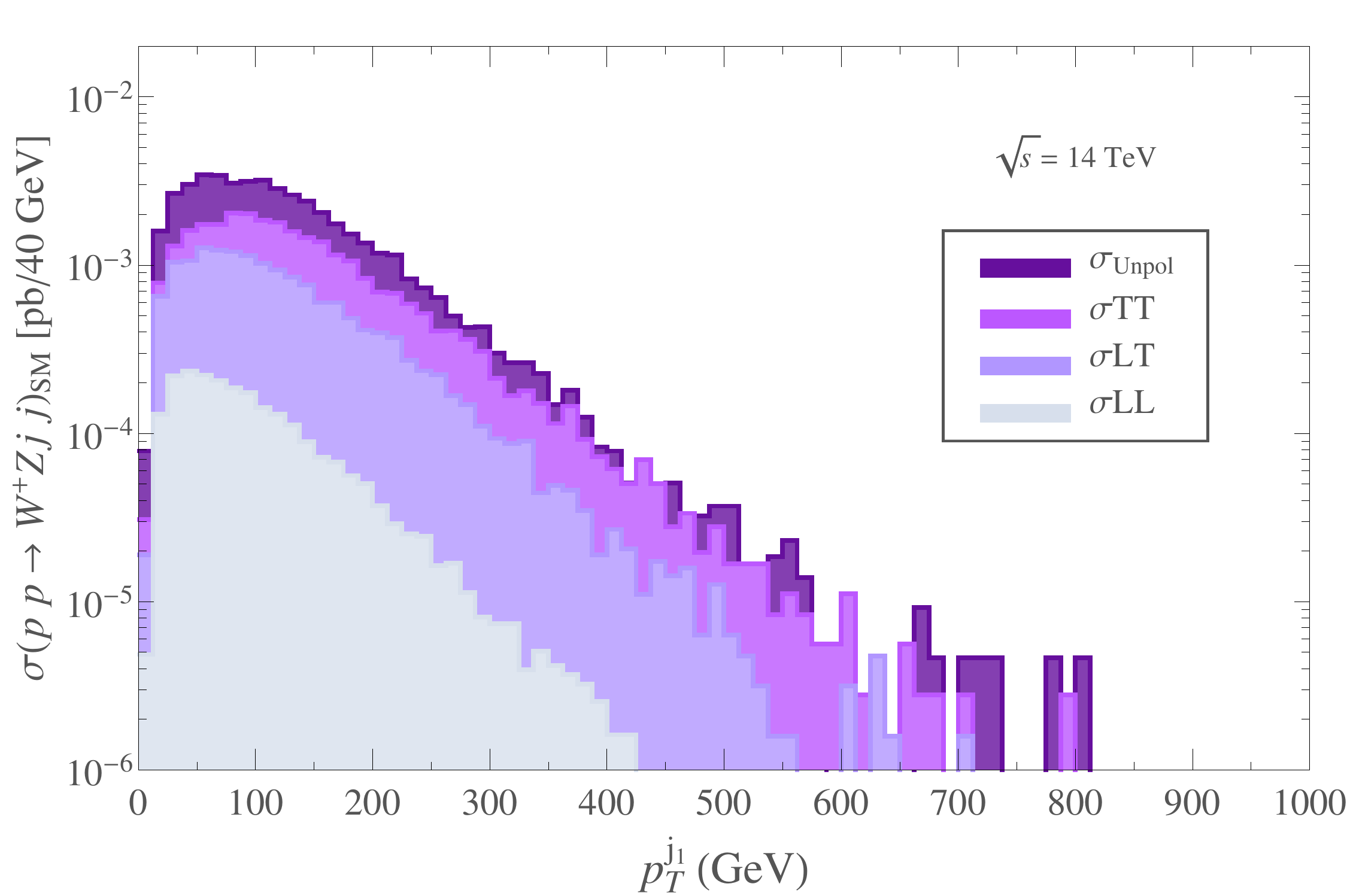}%
\hfill\includegraphics[width=.49\textwidth]{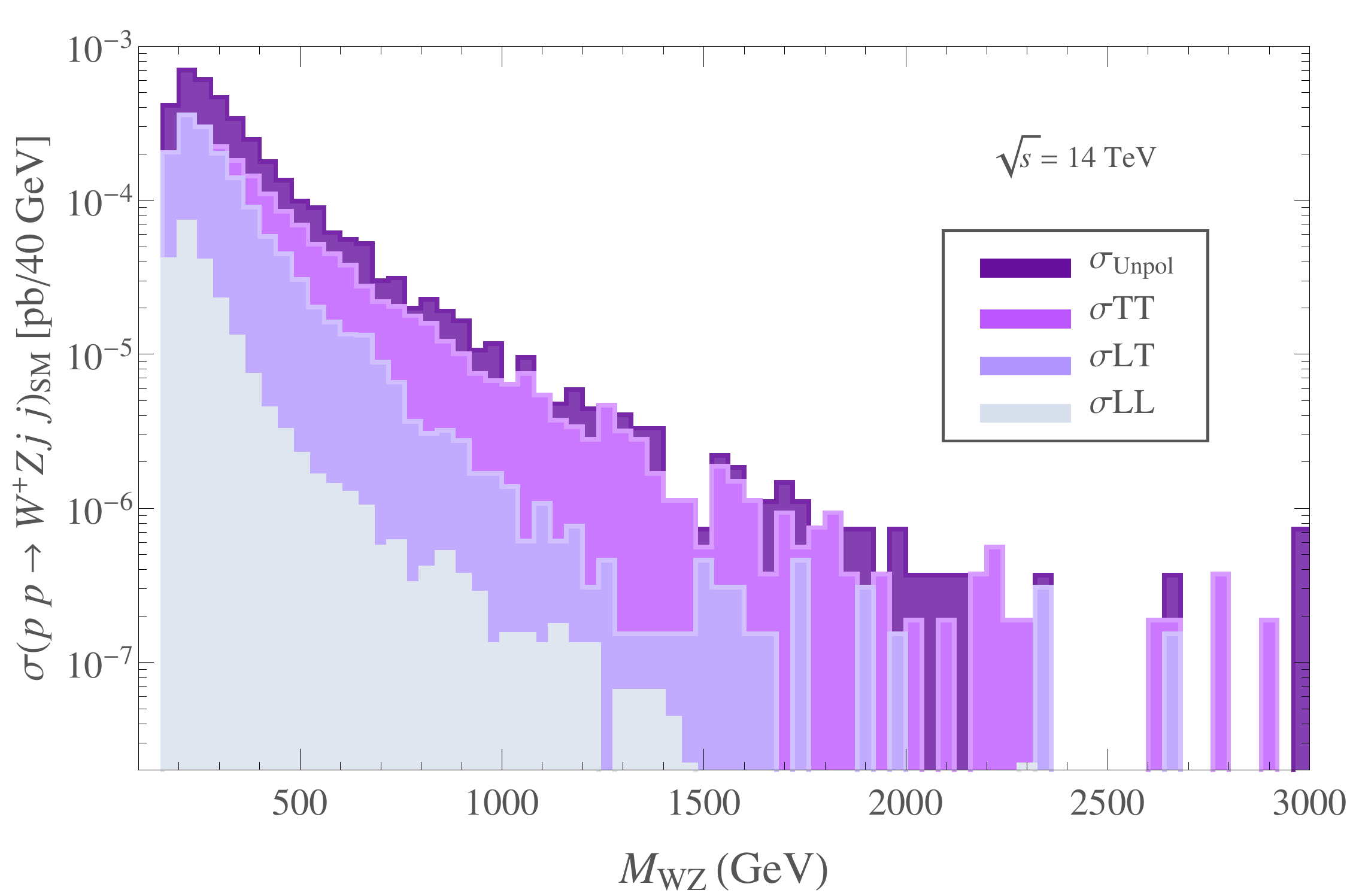}%
\hfill\null
\caption{\small Pure SM-EW background. Left, invariant mass of $W^+Z$. Right, transverse momentum of the most energetic jet. Cuts: $\lvert\eta_{j_1,j_2}\rvert<5$, $\eta_{j_1}\cdot\eta_{j_2}<0$, $\lvert\eta_{W,Z}\rvert<2$. Polarizations are separated. Note the suppression of longitudinal polarization in the SM.}\label{ProdSMfig}
\end{figure}

For each BP, we generate two runs. One, with $W^+Zjj$ as final state. The other one, including the leptonic decays $W^+\to l^+\nu$, $Z\to l^+l^-$. In all the cases, the following cuts are set over the final 2-jets: $2<\lvert\eta_{j_1,j_2}\rvert<5$, $\eta_{j_1}\cdot\eta_{j_2}<0$, $p_T^{j_1,j_2}>20\,{\rm GeV}$ and $M_{jj}>500\,{\rm GeV}$. For the $W^+Zjj$ final state run, an additional cut $\lvert\eta_{W,Z}\rvert < 2$ is used. For the leptonic decay run, we set the additional cuts $M_Z-10\,{\rm GeV} < M_{\ell^+_Z \ell^-_Z} < M_Z+10\,{\rm GeV}$, $M^T_{WZ}\equiv M^T_{\ell\ell\ell\nu}>500\,{\rm GeV}$,  $\slashed{p}_T>75\,{\rm GeV}$ and $p_T^\ell>100\,{\rm GeV}$. On top of the BSM signals, we have computed two SM backgrounds: pure SM-EW background (shown in Fig.~\ref{ProdSMfig}), $q_1q_2\to q_3q_4W^+Z$ scattering at order $\mO(\alpha^2)$. And mixed SM-QCDEW, at order $\mO(\alpha\alpha_S)$.

\begin{figure}[th]
\null%
\hfill\includegraphics[width=.49\textwidth]{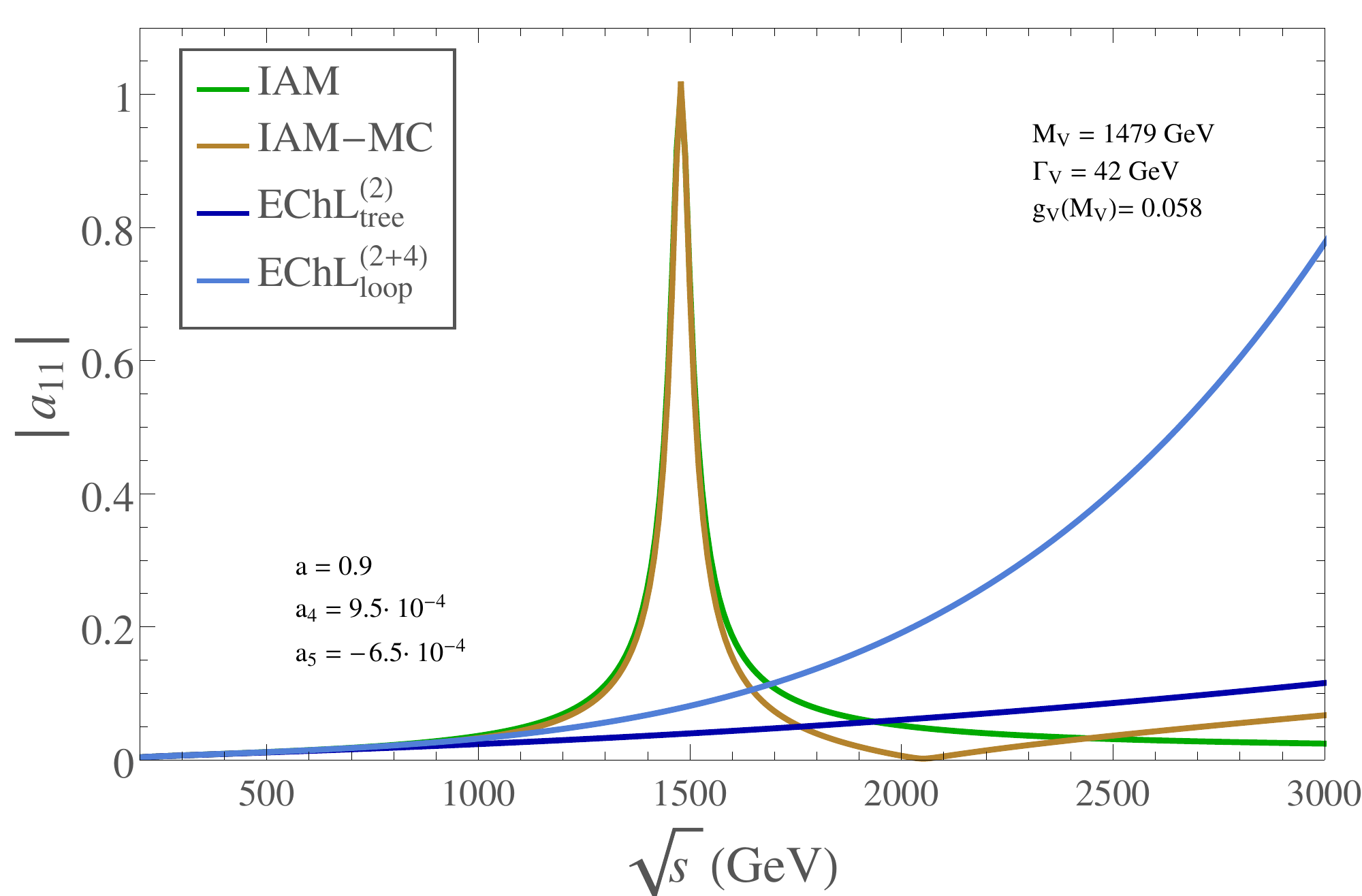}%
\hfill\includegraphics[width=.49\textwidth]{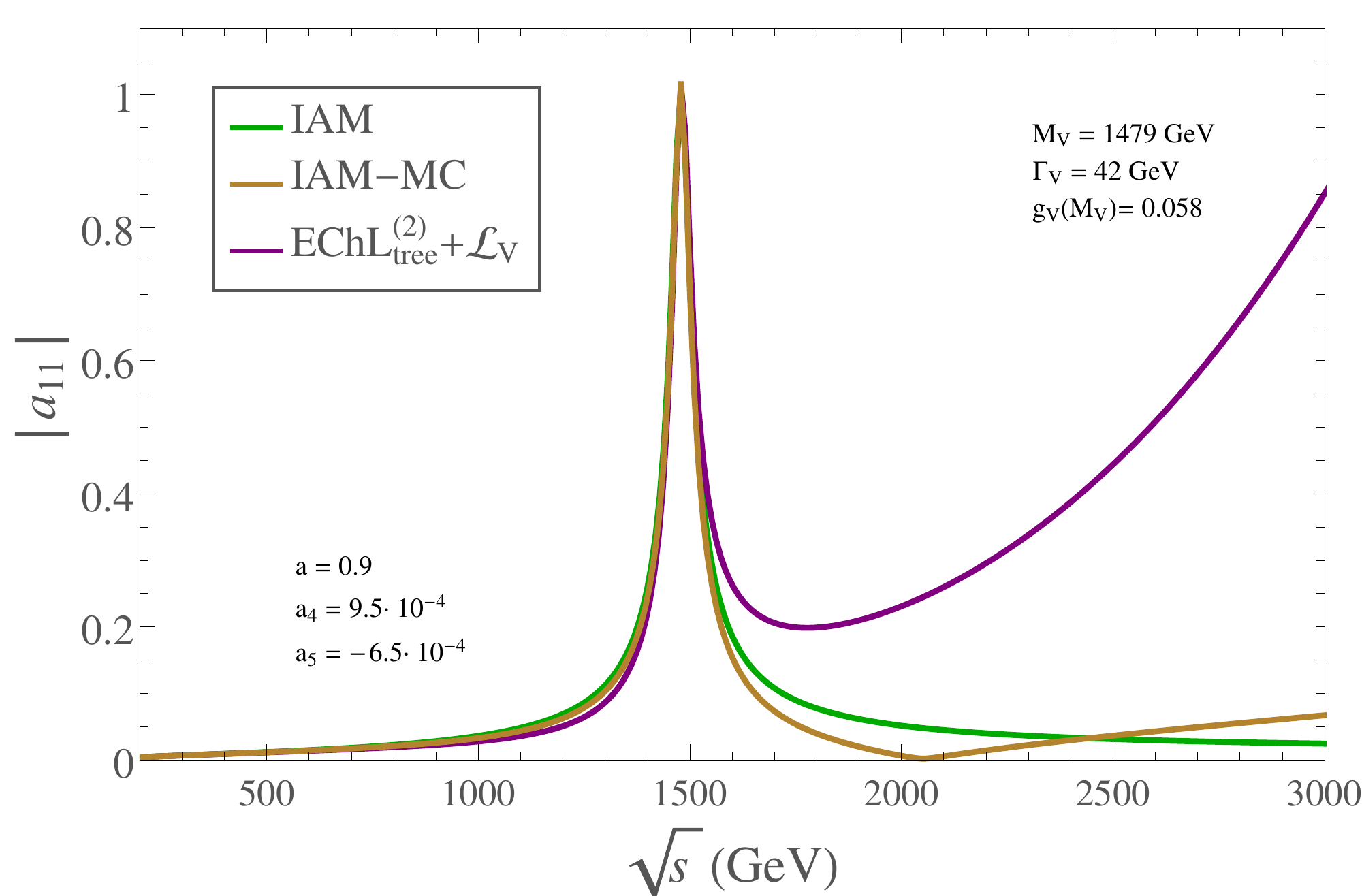}%
\hfill\null
\caption{\small Absolute value of the isovector-vector partial wave, $a_{11}$ ($IJ=11$) for BP1' (see table~\ref{BPtable}). ${\rm EChL}_{\rm tree}^{(2)}$ and ${\rm EChL}_{\rm loop}^{(2+4)}$, perturbative (non-unitarized) LO and NLO computation with the EWChL; IAM, unitarized partial wave; ${\rm EChL}_{\rm tree}^{(2)}+\mL_V$, perturbative LO computation in the EChL + Proca Lagrangian (constant $g_V$); IAM-MC, the MadGraph~5 model developed in this work.%
}\label{t11compar}
\end{figure}

\begin{figure}[th]
\null%
\hfill\includegraphics[width=.49\textwidth]{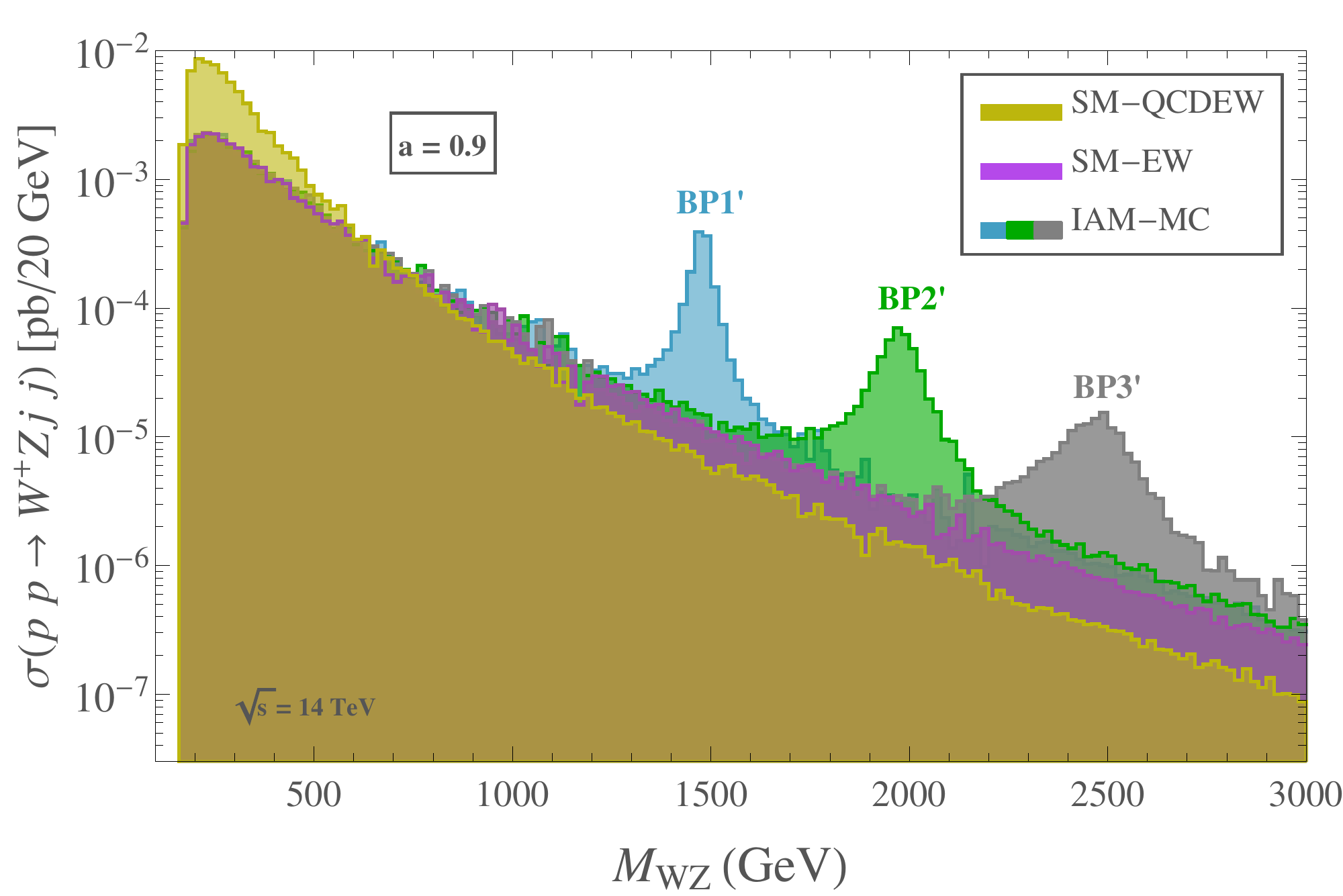}%
\hfill\includegraphics[width=.49\textwidth]{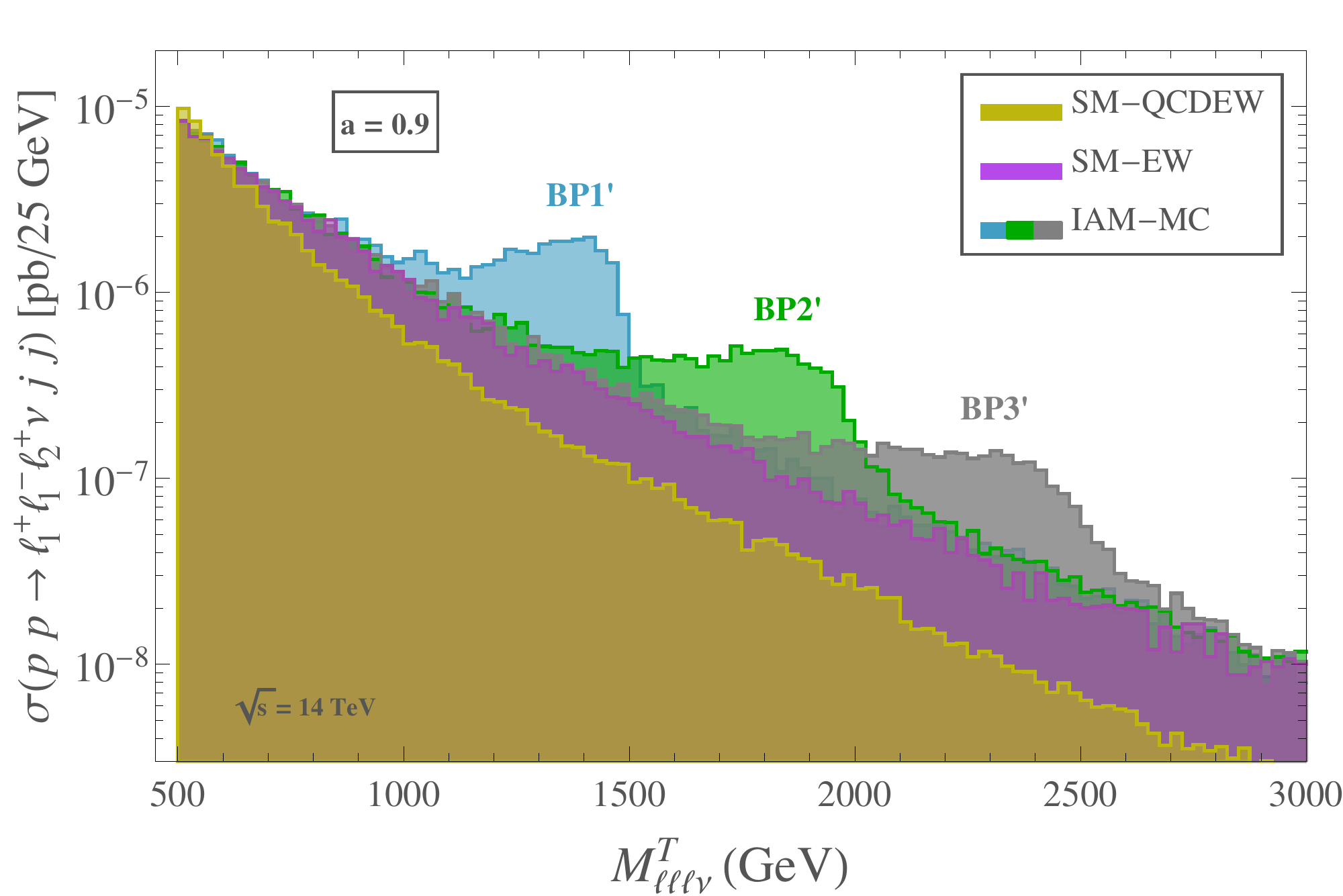}%
\hfill\null
\caption{\small BP1' (see table~\ref{BPtable}). $W^+Zjj$ in final state (left) vs. leptonic decay (right).}\label{BPa9}
\end{figure}

\begin{figure}[th]
\begin{center}
\includegraphics[width=.49\textwidth]{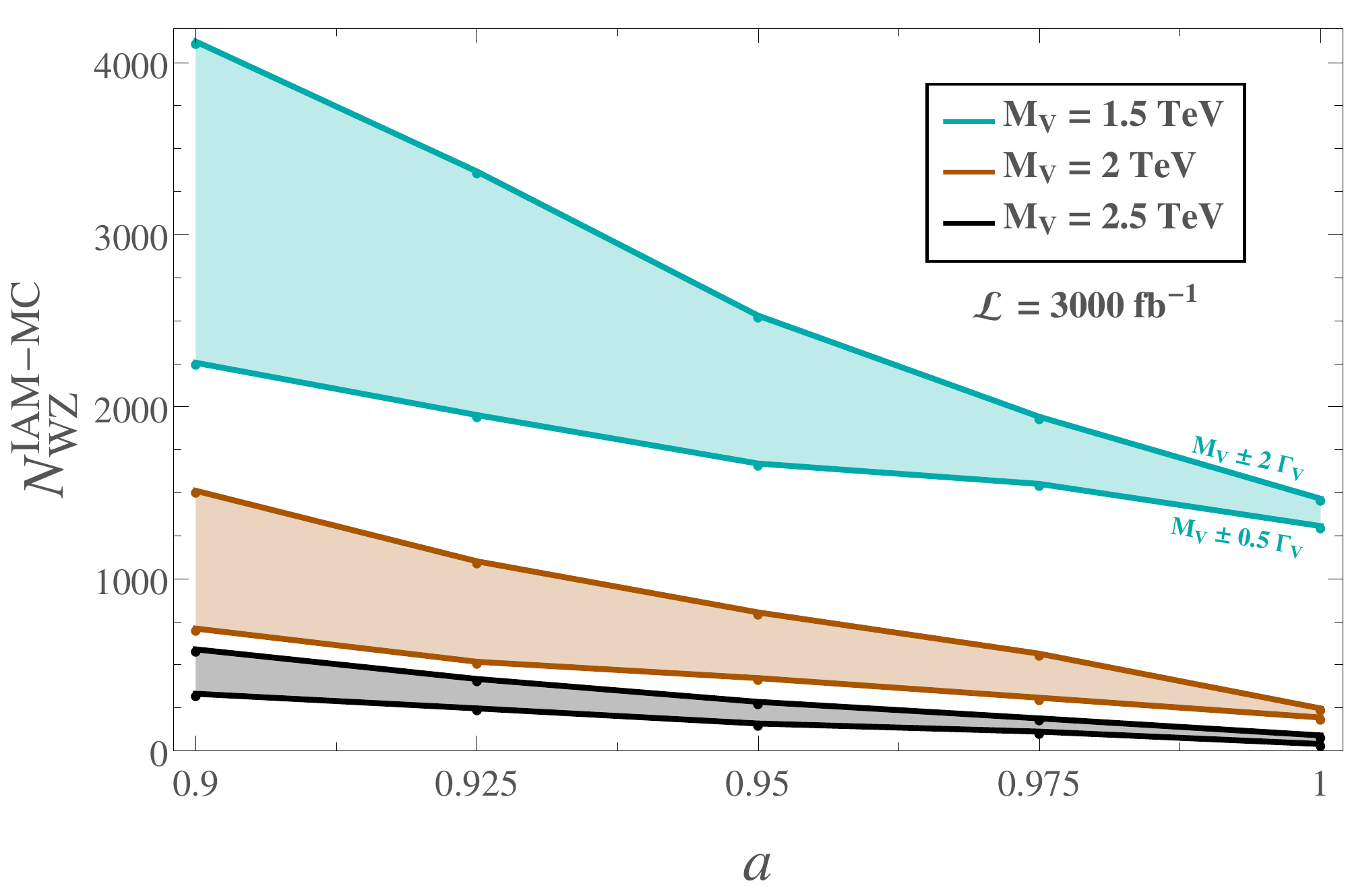}
\includegraphics[width=.48\textwidth]{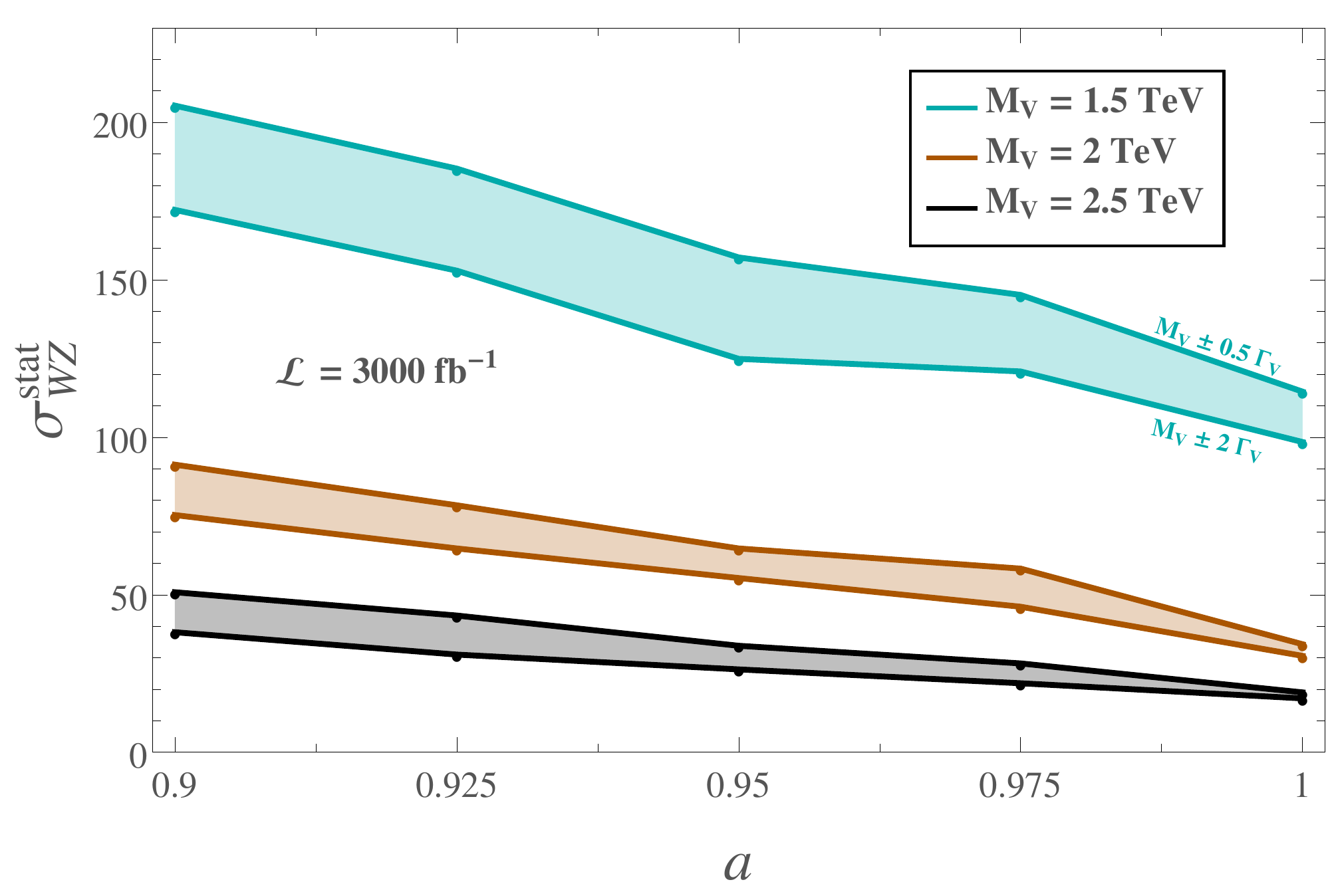}
\caption{Predictions for the number of events, ${\rm N}^{\rm IAM-MC}_{WZ}$ (left panel), and the statistical significance, $\significance_{WZ}$ (right panel), as a function of the parameter $a$ for $\mL=3000~{\rm fb}^{-1}$. The marked points correspond to our selected BPs in Fig.~\ref{BPfig}. The two lines for each mass are computed by summing events within $\pm 0.5\,\Gamma_V $ and $\pm 2\,\Gamma_V$, respectively.}\label{figevents}
\end{center}
\end{figure}

Finally, we make a prediction on the number of events expectable at $14\,{\rm TeV}$ for different LHC luminosities (see table~\ref{tablasigmaslep} and Fig.~\ref{figevents}). The results for the cross sections are shown in Fig.~\ref{BPa9}. The results for the relevant partial wave $a_{11}$ are shown in Fig.~\ref{t11compar}. For the statistical significance, we are using the standard expression $\significance_\ell=S_\ell/\sqrt{B_\ell}$, $S_\ell=N^{\rm IAM-MC} - N^{\rm SM}$, $B_\ell=N^{\rm SM}$, $N^i=N(pp\to l_1^+l_1^-l_2^+\slashed{p}_T jj)^i$, and the following ranges of $M_{lll\nu}^T$:
\begin{align}
&{\rm BP1:}~1325-1450~{\rm GeV}\,, && {\rm BP2:}~1875-2025~{\rm GeV}\,, && {\rm BP3:}~2300-2425~{\rm GeV}\,,\nn \\
&{\rm BP1':}~1250-1475~{\rm GeV}\,, && {\rm BP2':}~1675-2000~{\rm GeV}\,, &&{\rm BP3':}~2050-2475~{\rm GeV}\,.
\label{MTintervals}
\end{align}
The cases with $a=1$ have smaller significances, and only the lightest resonances $M_V=1.5\,{\rm TeV}$ (BP1) could be seen at $\sim 3\sigma$ and the highest luminosity ($3000\,{\rm fb}^{-1}$). Note that the cells without data means \emph{lack of statistics}. The cases with heavier $M_V\sim 2.5\,{\rm TeV}$ seem difficult to observe, due to poor statistics in the leptonic channels. Only BP3' obtain a significance $>2\sigma$ (for $3000\,{\rm fb}^{-1}$). Hence, semileptonic and fully hadronic channels look necessary to improve these significances. On the other case, the largest significances are obtained for $a=0.9$ and the lightest resonances, which corresponds to our BP1'. Significances $\sim 2.8\sigma$, $5.1\sigma$ and $8.9\sigma$ are predicted for LHC luminosities $\mL=300\,{\rm fb}^{-1}$, $1000\,{\rm fb}^{-1}$ and $3000\,{\rm fb}^{-1}$, respectively.

\begin{table}[th]
\begin{center}
\begin{tabular}{cc|c|c|c|c|c|c|}
\hhline{~~|-|-|-|-|-|-|}
& & \cellcolor{gray! 50}BP1 & \cellcolor{gray! 50}BP2 &\cellcolor{gray! 50} BP3 & \cellcolor{gray! 50}BP1' & \cellcolor{gray! 50}BP2' & \cellcolor{gray! 50}BP3' \\
\hhline{~|-|-|-|-|-|-|-|}
\multicolumn{1}{c|}{\multirow{3}{*}{\begin{sideways}$\mL=300\,{\rm fb}^{-1}$\end{sideways}}}&\cellcolor{gray! 15} ${\rm N}^{\rm IAM-MC}_{\ell}$ & 2 & 0.5 & 0.1 & 5 & 2 & 0.7 \\[0.5ex]
\hhline{~|-|-|-|-|-|-|-|}
\multicolumn{1}{c|}{} & \cellcolor{gray! 15}${\rm N}^{\rm SM}_{\ell}$ & 1 & 0.4 & 0.1 & 2 & 0.6 & 0.3 \\[0.5ex]
\hhline{~|-|-|-|-|-|-|-|}
\multicolumn{1}{c|}{} &\cellcolor{gray! 15} $\significance_{\ell}$ & 0.9 & - &- & 2.8 & 1.4 &- \\[0.5ex]
\hhline{~|-|-|-|-|-|-|-|}\\[-4ex]\hhline{~|-|-|-|-|-|-|-|}
\multicolumn{1}{c|}{\multirow{3}{*}{\begin{sideways}$\mL=1000\,{\rm fb}^{-1}$\end{sideways}}} &\cellcolor{gray! 15}${\rm N}^{\rm IAM-MC}_{\ell}$ &7 & 2 & 0.4 & 18 & 5 & 2 \\[0.5ex]
\hhline{~|-|-|-|-|-|-|-|}
\multicolumn{1}{c|}{}&\cellcolor{gray! 15} ${\rm N}^{\rm SM}_{\ell}$ & 4 & 1 & 0.3 & 6 & 2 & 1 \\[0.5ex]
\hhline{~|-|-|-|-|-|-|-|}
\multicolumn{1}{c|}{}& \cellcolor{gray! 15}$\significance_{\ell}$ & 1.6 & 0.3 & - & 5.1 & 2.5 & 1.4 \\[0.5ex]
\hhline{~|-|-|-|-|-|-|-|}\\[-4ex]\hhline{~|-|-|-|-|-|-|-|}
\multicolumn{1}{c|}{\multirow{3}{*}{\begin{sideways}$\mL=3000\,{\rm fb}^{-1}$\end{sideways}}} &\cellcolor{gray! 15} ${\rm N}^{\rm IAM-MC}_{\ell}$ & 22 & 5 & 1 & 53 & 16 & 7 \\[0.5ex]
\hhline{~|-|-|-|-|-|-|-|}
\multicolumn{1}{c|}{}& \cellcolor{gray! 15}${\rm N}^{\rm SM}_{\ell}$ & 12 & 4 & 1 & 17 & 6 & 3 \\[0.5ex]
\hhline{~|-|-|-|-|-|-|-|}
\multicolumn{1}{c|}{}& \cellcolor{gray! 15}$\significance_{\ell}$ & 2.7& 0.6 & 0.3 & 8.9 & 4.4 & 2.4 \\[0.5ex]
\hhline{~|-|-|-|-|-|-|-|}
\end{tabular}\caption{Predicted number of $pp\to l_1^+l_1^-l_2^+\nu jj$ events of the IAM-MC, ${\rm N}^{\rm IAM-MC}_l$, and of the SM background (EW+QCDEW), ${\rm N}^{\rm SM}_l$, at $14\,{\rm TeV}$, for different LHC luminosities: $\mL=300\,{\rm fb}^{-1}$, $\mL=1000\,{\rm fb}^{-1}$ and $\mL=3000 ~{\rm fb}^{-1}$. We also present the corresponding statistical significances, $\significance_\ell$.}\label{tablasigmaslep}
\end{center}
\end{table}

\section{Conclusions}
In this work, we have developed a MadGraph~5 model for strongly interacting Vector Boson Scattering in the isovector channel ($IJ=11$), by means of the Inverse Amplitude Method (IAM) and an effective Proca Lagrangian. The process $pp\to W^+Zjj$ via VBS, with a fully leptonic decay $W^+\to l^+\nu$, $Z\to l^+l^-$, has been studied.

We selected 6 benchmark points, with $M_V=1.5$, $2$, $2.5\,{\rm TeV}$ and $a=0.9$, $1$. We have selected our BPs to make a first scan of the parameter space of the chiral parameters $a\in (0.9,1)$, $b=a^2$ and $a_4,\,a_5\in (10^{-4},10^{-3})$. For the sake of completeness, we have included on Fig.~\ref{BPfig} additional intermediate points to show the dependence of $M_V$, $\Gamma_V$ and $g_V(M_V^2)$ on the chiral parameters. For each benchmark point, a MadGraph~5 Monte Carlo model has been developed and run, both with and without leptonic decay. For the sake of brevity, only BP1' is reproduced here. However, the full analysis can be found on Ref.~\cite{Delgado:2017cls}.

Finally, we have included a prediction of number of events for several LHC luminosities at $14\,{\rm TeV}$ (see table~\ref{tablasigmaslep}). As discussed on section~\ref{secresults}, semileptonic and hadronic studies seem necessary in order to improve the sensitivity of LHC Run-II to the BSM effects in this process. Besides the leptonic channels considered here, a discussion on the semileptonic and hadronic channels can be found in~\cite{Delgado:2017cls}.

\section{Acknowledgements}
We thank P.~Arnan for providing us with the FORTRAN code to localize the IAM resonances and for his help at the early stages of this work. A.D. thanks F.J. Llanes-Estrada for previous collaboration. R.L.D. and J.J.S.C. thank Corinne Goy for useful discussions. This work is supported by the European Union through the ITN ELUSIVES H2020-MSCA-ITN-2015//674896, the RISE INVISIBLESPLUS H2020-MSCA-RISE-2015//690575 and the STSM Grant from COST Action CA16108, by the Spanish MINECO through the projects FPA2013-46570-C2-1-P, FPA2014-53375-C2-1-P, FPA2016-75654-C2-1-P, FPA2016-76005-C2-1-P, FPA2016-78645-P(MINECO/ FEDER, EU), by the Spanish Consolider-Ingenio 2010 Programme CPAN (CSD2007-00042) and by the Spanish MINECO's ``Centro de Excelencia Severo Ochoa'' Programme under grants SEV-2012-0249 and SEV-2016-0597 and the ``Mar\'ia de Maeztu'' Programme under grant MDM-2014-0369. R.L.D is supported by the MINECO project FIS2013-41716-P, FPA2016-75654-C2-1-P and the ``Ram\'on Areces'' Foundation. We also acknowledge 8000 hours of computer time granted at a small departmental cluster at the UCM.


\begin{thebibliography}{99}

\bibitem{Delgado:2017cls}
  R.~L.~Delgado, A.~Dobado, D.~Espriu, C.~Garcia-Garcia, M.~J.~Herrero, X.~Marcano and J.~J.~Sanz-Cillero,
  JHEP {\bf 1711} (2017) 098.

\bibitem{Aad:2012tfa}
  G.~Aad {\it et al.} [ATLAS Collaboration],
  Phys.\ Lett.\ B {\bf 716} (2012) 1.

\bibitem{Chatrchyan:2012xdj}
  S.~Chatrchyan {\it et al.} [CMS Collaboration],
  Phys.\ Lett.\ B {\bf 716} (2012) 30.

\bibitem{Khachatryan:2014jba}
  V.~Khachatryan {\it et al.} [CMS Collaboration],
  Eur.\ Phys.\ J.\ C {\bf 75} (2015) no.5, 212.

\bibitem{Aaboud:2017eta}
  M.~Aaboud {\it et al.} [ATLAS Collaboration],
  Phys.\ Lett.\ B {\bf 777} (2018) 91.

\bibitem{Sirunyan:2017acf}
  A.~M.~Sirunyan {\it et al.} [CMS Collaboration],
  Phys.\ Rev.\ D {\bf 97} (2018) no.7, 072006.

\bibitem{Appelquist:1980vg}
  T.~Appelquist and C.~W.~Bernard,
  Phys.\ Rev.\ D {\bf 22} (1980) 200.

\bibitem{Longhitano:1980iz}
  A.~C.~Longhitano,
  Phys.\ Rev.\ D {\bf 22} (1980) 1166.

\bibitem{Longhitano:1980tm}
  A.~C.~Longhitano,
  Nucl.\ Phys.\ B {\bf 188} (1981) 118.

\bibitem{Chanowitz:1985hj}
  M.~S.~Chanowitz and M.~K.~Gaillard,
  Nucl.\ Phys.\ B {\bf 261} (1985) 379.

\bibitem{Cheyette:1987jf}
  O.~Cheyette and M.~K.~Gaillard,
  Phys.\ Lett.\ B {\bf 197} (1987) 205.

\bibitem{Dobado:1989ax}
  A.~Dobado and M.~J.~Herrero,
  Phys.\ Lett.\ B {\bf 228} (1989) 495.
  
\bibitem{Dobado:1989ue}
  A.~Dobado and M.~J.~Herrero,
  Phys.\ Lett.\ B {\bf 233} (1989) 505.

\bibitem{Weinberg:1978kz}
  S.~Weinberg,
  Physica A {\bf 96} (1979) no.1-2,  327.

\bibitem{Gasser:1984gg}
  J.~Gasser and H.~Leutwyler,
  Nucl.\ Phys.\ B {\bf 250} (1985) 465.

\bibitem{Gasser:1983yg}
  J.~Gasser and H.~Leutwyler,
  Annals Phys.\  {\bf 158} (1984) 142.

\bibitem{Dobado:1990zh} 
  A.~Dobado, D.~Espriu and M.~J.~Herrero,
  Phys.\ Lett.\ B {\bf 255}, 405 (1991).

\bibitem{Espriu:1991vm}
  D.~Espriu and M.~J.~Herrero,
  Nucl.\ Phys.\ B {\bf 373} (1992) 117.
  
\bibitem{Dobado:1990jy}
  A.~Dobado, M.~J.~Herrero and J.~Terron,
  Z.\ Phys.\ C {\bf 50} (1991) 205.

\bibitem{Dobado:1990am}
  A.~Dobado, M.~J.~Herrero and J.~Terron,
  Z.\ Phys.\ C {\bf 50} (1991) 465.

\bibitem{Dobado:1995qy}
  A.~Dobado, M.~J.~Herrero, J.~R.~Pelaez, E.~Ruiz Morales and M.~T.~Urdiales,
  Phys.\ Lett.\ B {\bf 352} (1995) 400.

\bibitem{Dobado:1999xb}
  A.~Dobado, M.~J.~Herrero, J.~R.~Pelaez and E.~Ruiz Morales,
  Phys.\ Rev.\ D {\bf 62} (2000) 055011.

\bibitem{Feruglio:1992wf} 
  F.~Feruglio,
  Int.\ J.\ Mod.\ Phys.\ A {\bf 8}, 4937 (1993).

\bibitem{Alonso:2012px}
  R.~Alonso, M.~B.~Gavela, L.~Merlo, S.~Rigolin and J.~Yepes,
  Phys.\ Lett.\ B {\bf 722} (2013) 330
   Erratum: [Phys.\ Lett.\ B {\bf 726} (2013) 926].

\bibitem{Buchalla:2013rka}
  G.~Buchalla, O.~Cat\`a and C.~Krause,
  Nucl.\ Phys.\ B {\bf 880} (2014) 552
   Erratum: [Nucl.\ Phys.\ B {\bf 913} (2016) 475].

\bibitem{Espriu:2012ih}
  D.~Espriu and B.~Yencho,
  Phys.\ Rev.\ D {\bf 87} (2013) no.5,  055017.

\bibitem{Delgado:2013loa}
  R.~L.~Delgado, A.~Dobado and F.~J.~Llanes-Estrada,
  J.\ Phys.\ G {\bf 41} (2014) 025002.

\bibitem{Delgado:2013hxa}
  R.~L.~Delgado, A.~Dobado and F.~J.~Llanes-Estrada,
  JHEP {\bf 1402} (2014) 121.

\bibitem{Brivio:2013pma}
  I.~Brivio, T.~Corbett, O.~J.~P.~\'Eboli, M.~B.~Gavela, J.~Gonzalez-Fraile, M.~C.~Gonzalez-Garcia, L.~Merlo and S.~Rigolin,
  JHEP {\bf 1403} (2014) 024.

\bibitem{Espriu:2013fia}
  D.~Espriu, F.~Mescia and B.~Yencho,
  Phys.\ Rev.\ D {\bf 88} (2013) 055002.

\bibitem{Espriu:2014jya}
  D.~Espriu and F.~Mescia,
  Phys.\ Rev.\ D {\bf 90} (2014) no.1,  015035.

\bibitem{Delgado:2014jda}
  R.~L.~Delgado, A.~Dobado, M.~J.~Herrero and J.~J.~Sanz-Cillero,
  JHEP {\bf 1407} (2014) 149.

\bibitem{Buchalla:2015qju}
  G.~Buchalla, O.~Cata, A.~Celis and C.~Krause,
  Eur.\ Phys.\ J.\ C {\bf 76} (2016) no.5, 233.

\bibitem{Arnan:2015csa}
  P.~Arnan, D.~Espriu and F.~Mescia,
  Phys.\ Rev.\ D {\bf 93} (2016) no.1, 015020.

\bibitem{Delgado:2015kxa}
  R.~L.~Delgado, A.~Dobado and F.~J.~Llanes-Estrada,
  Phys.\ Rev.\ D {\bf 91} (2015) no.7,  075017.
  
\bibitem{Falkowski:2013dza}
  A.~Falkowski, F.~Riva and A.~Urbano,
  JHEP {\bf 1311} (2013) 111.

\bibitem{Aad:2014zda}
  G.~Aad {\it et al.} [ATLAS Collaboration],
  Phys.\ Rev.\ Lett.\  {\bf 113} (2014) no.14,  141803.

\bibitem{ATLAS:2014yka}
  The ATLAS collaboration [ATLAS Collaboration],
  ATLAS-CONF-2014-009.

\bibitem{Fabbrichesi:2015hsa}
  M.~Fabbrichesi, M.~Pinamonti, A.~Tonero and A.~Urbano,
  Phys.\ Rev.\ D {\bf 93} (2016) no.1,  015004.

\bibitem{Aaboud:2016uuk}
  M.~Aaboud {\it et al.} [ATLAS Collaboration],
  Phys.\ Rev.\ D {\bf 95} (2017) no.3,  032001.
  
\bibitem{deBlas:2018tjm}
  J.~de Blas, O.~Eberhardt and C.~Krause,
  JHEP {\bf 1807} (2018) 048.

\bibitem{Arnold:2008rz}
  K.~Arnold {\it et al.},
  Comput.\ Phys.\ Commun.\  {\bf 180} (2009) 1661.
  
\bibitem{Kilian:2014zja}
  W.~Kilian, T.~Ohl, J.~Reuter and M.~Sekulla,
  Phys.\ Rev.\ D {\bf 91} (2015) 096007.

\bibitem{Contino:2013kra}
  R.~Contino, M.~Ghezzi, C.~Grojean, M.~Muhlleitner and M.~Spira,
  JHEP {\bf 1307} (2013) 035.

\bibitem{Alloul:2013naa}
  A.~Alloul, B.~Fuks and V.~Sanz,
  JHEP {\bf 1404} (2014) 110.

\bibitem{Alwall:2014hca}
  J.~Alwall {\it et al.},
  JHEP {\bf 1407} (2014) 079.

\bibitem{Frederix:2018nkq}
  R.~Frederix, S.~Frixione, V.~Hirschi, D.~Pagani, H.-S.~Shao and M.~Zaro,
  JHEP {\bf 1807} (2018) 185.
  
\bibitem{Alloul:2013bka}
  A.~Alloul, N.~D.~Christensen, C.~Degrande, C.~Duhr and B.~Fuks,
  Comput.\ Phys.\ Commun.\  {\bf 185} (2014) 2250.

\bibitem{Degrande:2011ua}
  C.~Degrande, C.~Duhr, B.~Fuks, D.~Grellscheid, O.~Mattelaer and T.~Reiter,
  Comput.\ Phys.\ Commun.\  {\bf 183} (2012) 1201.

\end{thebibliography}
\end{document}